\documentclass[12pt,epsf]{article}
\usepackage{graphicx}

\begin{document}

\sloppy
\begin{flushright}{SIT-HEP/TM-22}
\end{flushright}
\vskip 1.5 truecm
\centerline{\large{\bf String production after angled brane
inflation}} 
\vskip .75 truecm
\centerline{\bf Tomohiro Matsuda
\footnote{matsuda@sit.ac.jp}}
\vskip .4 truecm
\centerline {\it Laboratory of Physics, Saitama Institute of
 Technology,}
\centerline {\it Fusaiji, Okabe-machi, Saitama 369-0293, 
Japan}
\vskip 1. truecm
\makeatletter
\@addtoreset{equation}{section}
\def\theequation{\thesection.\arabic{equation}}
\makeatother
\vskip 1. truecm

\begin{abstract}
\hspace*{\parindent}
We describe string production after angled brane inflation.
First, we point out that there was a discrepancy in previous discussions.
The expected tension of the cosmic string calculated from the
 four-dimensional effective Lagrangian did not match the one obtained
 in the brane analysis.  
In the previous analysis, the cosmic string is assumed
to correspond to the lower-dimensional daughter brane,
which wraps the same compactified space as the original mother brane.
In this case, however, the tension of the daughter brane cannot depend
on the angle ($\theta$).
On the other hand, from the analysis of the effective Lagrangian for
tachyon condensation, it is easy to see that the tension of the 
cosmic string must be proportional to $\theta$, when $\theta \ll 1$.
This is an obvious discrepancy that must be explained by
 consideration of the explicit brane dynamics.
In this paper, we will solve this problem by introducing a simple idea.
We calculate the tension of the string in the two cases, which matches
 precisely. 
The cosmological constraint for angled inflation is relaxed, 
because the expected tension of the cosmic string becomes smaller than
the one obtained in previous arguments, by a factor of $\theta$. 
\end{abstract}

\newpage
\section{Introduction}
String theory is perhaps the most promising scenario where quantum
gravity is included by the requirement of additional dimensions and
supersymmetry.
Models with more than four dimensions are interesting, since
the idea of the large extra dimension\cite{Extra_1} may solve or weaken
the hierarchy problem.\footnote{Denoting the volume of the
$n$-dimensional compact space by $V_n$, 
the observed Planck mass is obtained by the relation $M_p^2=M^{n+2}_{*}V_n$,
where $M_{*}$ denotes the fundamental scale of gravity.}
In scenarios of large extra dimensions, the standard model fields
are expected to be localized on a wall-like structure, while the graviton
propagates in the bulk. 
In the context of string theory, the most natural embedding of
this picture would be realized by a brane construction.
These models are of course very interesting from the cosmological viewpoint.
Inflation with such a situation is still an interesting
problem\cite{low_inflation, matsuda_nontach}.
A model of defect inflation with large extra dimension is discussed in
\cite{matsuda_defectinfla}.
Other cosmological issues,  in which defects play important roles, such
as baryogenesis with a low fundamental 
scale are discussed in \cite{low_baryo, Defect-baryo-largeextra,
Defect-baryo-4D}. 
We think cosmological considerations with nonstatic brane
configurations, such as brane defects or brane Q balls\cite{BraneQball,
MQCD_defect_matsuda} are very important, since
we expect that future cosmological observations will reveal
the cosmological evolution of the brane Universe.
If one wants to know what kinds of brane defects are
allowed to affect our present Universe, one should know how they were
formed.  
In the original brane inflation scenario\cite{brane-inflation0},
the inflationary expansion is driven by the potential between D-branes
and anti-D-branes evolving in the bulk space of the compactified
dimensions.
Then the brane inflationary scenario for branes at a fixed angle is studied
in ref.\cite{angled-inflation}, where the slow-roll condition is 
improved by a small angle.
The end of brane inflation is induced by the brane collision where brane
annihilation (or recombination) proceeds through tachyon
condensation\cite{tachyon0}. 
During brane inflation, the tachyon is trapped in the false vacuum, which
may result in the formation of lower-dimensional branes after brane inflation.
The production of cosmological defects during this process is discussed
in ref.\cite{angled-inflation, angled-defect}, where it is concluded
that cosmic strings 
are copiously produced in this scenario, but the formation of domain
walls is seriously suppressed.
Ref.\cite{Majumdar_Davis}, however, indicates that all kinds 
of defects would be produced and the conventional problems of cosmic
domain walls and monopoles should arise.
Later, in refs.\cite{D-brane-strings, Halyo, BDKP-FI}, brane
production was reexamined 
and the conclusions were different from \cite{angled-defect} and
\cite{Majumdar_Davis}. \footnote{The actual phenomenological possibilities of 
cosmological brane defects in the brane world models are not confined within
the above arguments. 
For example, in ref.\cite{Defect-baryo-largeextra}, the enhancements of
baryon number violation and baryogenesis that are mediated by
cosmological defects are discussed for models with localized matter
fields.
Moreover, in ref.\cite{Alice-string} and later in
refs.\cite{MQCD_defect_matsuda, incidental}, it  
is argued that the fields that parameterize the position of the branes can
fluctuate in four-dimensional space-time, and the fluctuation forms a
cosmological defect.
In ref.\cite{Alice-string}, the position of a brane in the 
fifth dimension is used to parameterize the Alice
string in the effective four-dimensional space-time, and it was shown that
the brane must smear in the core so that it resolves the anticipated
singularity.  
Then in refs.\cite{MQCD_defect_matsuda} and \cite{incidental}, it is
shown that the relative position of 
the branes in higher-dimensional bulk space could fluctuate in the
four-dimensional space-time and could parameterize cosmological
defects such as strings, domain walls, and Q balls.}

In this paper, we will examine the tension of the cosmic
strings  produced after angled brane inflation.
Here we summarize the naive expectations and known facts.
\begin{itemize}
\item The model of angled brane inflation is expected to be a model of 
D-term inflation in the four-dimensional effective Lagrangian.
\item A cosmic string that is produced after D-term inflation is examined in
      ref.\cite{D-term-string}, and is shown to satisfy the BPS
      conditions in a generic situation. The exact form of the tension
      of the D-term string is also calculated.
\end{itemize}
Thus, one would expect that the cosmic string that is produced after
angled brane inflation is the D-term string that satisfies the BPS
condition.
If the above expectations are correct, the tension of the cosmic string
should depend on the angle. 
On the other hand, however, the previous arguments on string
formation after angled brane inflation did not support
this expectation.
For example, in refs.\cite{angled-defect, D-brane-strings}, it is
discussed that; 
\begin{itemize}
\item The produced daughter brane must wrap around the same compactified
      space as the original mother brane. Thus, the calculated tension of
      the daughter brane does not depend on the angle.
\end{itemize}
This result does not support the above expectation.
Moreover, one can calculate the tension of the cosmic string from the
effective Lagrangian for tachyon condensation.
The tension obtained from the tachyon effective Lagrangian does depend
on the angle, 
which again does not support the argument in refs.\cite{angled-defect,
D-brane-strings}. 

In this paper, we will show that one can easily understand why there was
such a discrepancy between the two calculations.
Our result is that the tension of the produced cosmic string does
depend on the angle, which is different from previous discussions.
The final state of the produced daughter brane does not wrap the same
compactified space as the original mother brane.
However, here we should note that our result does not contradict to
the previous arguments.
We will consider the details of the recombination process of the 
original branes, and find that the daughter brane that wraps the same
compactified space as the original mother brane would appear as the precedent
state of the final brane configuration.
The precedent brane configuration is then deformed into the final
state, where the daughter brane is extended between recombined branes.
The tension of the daughter brane now depends on the angle, which
is consistent with the expectation from the effective Lagrangian.

In Sec. 2, we begin with a short review of an angled brane inflation
model and the cosmic string production after
inflation\cite{angled-defect}.
From the viewpoint of the effective Lagrangian, it is easy to see that
the tension of the strings produced  must depend on the angle $\theta$.
In the previous arguments\cite{angled-defect}, however, it was
assumed that the daughter brane wraps the same compactified
space as the mother brane.
On the basis of this assumption, the tension of the cosmic
string was calculated and did not depend on $\theta$.
However, if one pays further attention to the process of recombination
of the mother branes, one can easily understand why the final state of
the brane produced does not wrap the same compactified space as the
mother brane.
The tension of the cosmic string does depend on the angle
$\theta$, which is of course consistent with the analysis in the
effective Lagrangian. 
The cosmological constraints are modified and will be relaxed by a
factor of $\theta$.

\section{String production after angled brane inflation}
Let us start with a short review of angled inflation.
Here we mainly follow ref.\cite{angled-defect} for later convenience.
In this section we consider string models that have 6 of the 9 space
dimensions compactified, and $D_p$-branes in 10-dimensional space-time.
Here the $(p-3)$ dimensions parallel to the $D_p$-branes are 
compactified with volume $V_{\|}$, and the $d$ dimensions orthogonal to
the $D_p$-branes are compactified with volume $V_{d}$.
The remaining 3 dimensions of the $D_p$-branes are not compactified and
they are the three spatial dimensions of the four-dimensional effective
Lagrangian. 
Here we denote the string scale by $M_s^2\equiv\alpha'^{-1}$.
Here we will assume that only $d_{\bot}$ dimensions of the orthogonal $d$
dimensions are large, and the volume of the compactified 6-dimensional
space is $V=V_{\|} V_{\bot}V_{d-d_{\bot}}$.
The volume $V_{d-d_{\bot}}$ is assumed to be
$V_{d-d_{\bot}}=(2\pi/M_s)^{d-d_{\bot}}$.
For simplicity, we will also assume that all the large extra dimensions
have an equal size $l_{\bot}=2\pi r_{\bot}$, which allows us to use 
$V_{\bot}=(l_{\bot})^{d_{\bot}}$ for the volume that is transverse to
the $D_p$-brane.
For the volume of the compactified space of the $D_p$-brane, we use
$V_{\|}$.

String models with branes at angles have already been discussed by many
authors.
Here we consider angled $D_p$-branes where each brane is wrapping 1-cycles
in a two dimensional torus that have radii $r_{\|}$ and $u r_{\|}$,
where $u$ is a constant of $u<1$.\footnote{For simplicity, we use 
$u=1$ in later calculations.}
Then the energy density of the brane is given by\cite{angled-defect}
\begin{equation}
E_{i} =\tau_4 l_{\|} \left[n_i^2 + (u m_i)^2\right]^{1/2},
\end{equation}
where the brane wraps a straight line in the ${n_i[a]+m_i[b]}$ homology
class, and $\tau_p = \frac{M_s^{p+1}}{(2\pi)^p g_s}$.
Let us consider two angled $D_p$ branes at a distance in the
direction orthogonal to the torus.
The angle $\theta$ is given by $\theta = \phi_1 -\phi_2$, where $\phi_i
=\tan^{-1}(um_i/n_i)$.
When the distance is large, the effective potential for angled
inflation is given by
\begin{equation}
\label{inflation-potential}
V(y) = V_0 -\frac{\beta M_s^{6-d_{\|}}\sin^2\theta/2 \tan \theta/2}
{4\pi y^{d_{\bot}-2}}
\end{equation}
for $d_\bot >2$.
Here $\beta$ is a constant, and $V_0$ is proportional to $\tan^2 \theta$.
The potential becomes logarithmic when $d_\bot =2$.

Cosmic strings would be produced due to tachyon condensation
during recombination after angled inflation.
Tachyon condensation generically allows the formation of
lower-dimensional branes.
The potential for the tachyon $T$ is\cite{angled-defect}
\begin{equation}
\label{tachyon-potential}
V_T \propto tr(TT^{\dag})(M_s^4 y^2 - 2\pi M_s^2 \theta),
\end{equation}
for $\theta \ll 1$.
From eq.(\ref{inflation-potential}) and eq.(\ref{tachyon-potential}),
one could easily understand the $\theta$-dependence of the cosmic
strings, $\mu \propto \theta$.
Thus, the calculation in the effective Lagrangian suggests that the
tension of the cosmic string must be proportional to $\theta$.

On the other hand, however, it is also suggested that the Kibble
mechanism does not happen in compactified dimensions, so that 
cosmic strings would be formed by the $D_{p-2}$ -branes wrapping the
same compactified cycles as the original $p$-branes.
The tension of such cosmic string is
$\mu=\frac{M_s^{p-1}V_{\|}}{(2\pi)^{p-2}g_s}$, which cannot depend on
$\theta$.

As we stated in the previous section, the two results are in
contradiction, which must be explained clearly from 
the brane dynamics during recombination.

\begin{figure}[ht]
 \begin{center}
\begin{picture}(300,100)(0,0)
\resizebox{10cm}{!}{\includegraphics{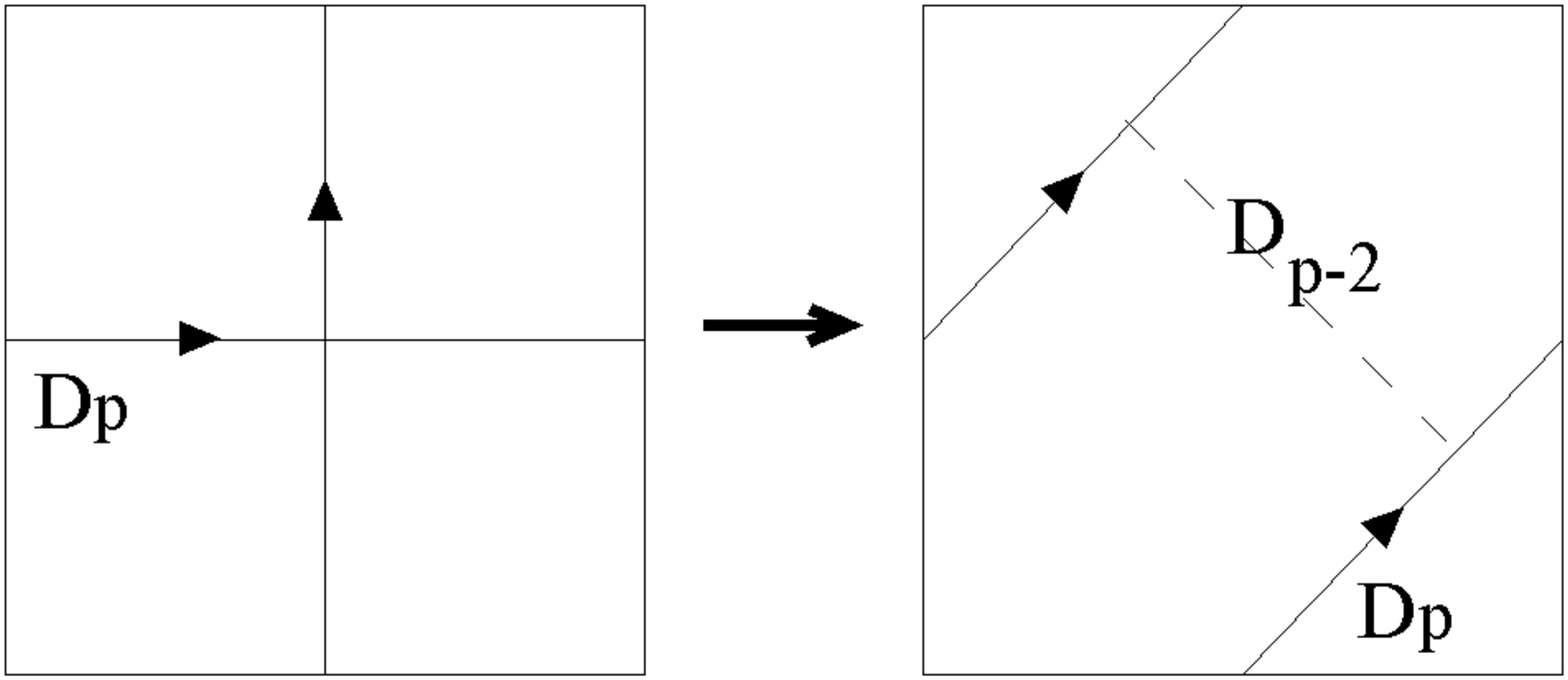}} 
\end{picture}
\caption{Recombination of the two-brane system on a two-dimensional
  torus. The dashed line is the $D_{p-2}$-brane that corresponds to the
  cosmic string in four-dimensional space-time.}  
\label{fig:basic}
 \end{center}
\end{figure}

First we will naively assume that the produced cosmic strings are the
$D_{p-2}$-branes that are stretched between final state of the
$D_p$-brane, as is depicted in fig.\ref{fig:basic}. 
Then the tension of the $D_{p-2}$-brane depends on the distance between
branes, which 
is proportional to $\theta$ for $\theta \ll 1$.
If the supersymmetry breakings induced by other source branes at a
distance are
negligible, the final state of the angled inflation is the BPS state
of a $D_p$-brane.
On the basis of our assumption that the cosmic string corresponds to
the stretched $D_{p-2}$-brane, the produced
cosmic string will also be a BPS state.
According to the discussion in ref.\cite{D-term-string}, the cosmic
string that satisfies the BPS conditions is the D-term string in the
effective Lagrangian.
Thus, our assumption is consistent with the expectation that the angled
inflation is described as a model of D-term inflation in the effective
Lagrangian.\footnote{An explicit 
calculation for the correspondence between the D-term string in the
effective Lagrangian and the extended $D$-brane is given in
ref.\cite{MQCD_defect_matsuda} for an inflation model due to Hanany-Witten
type brane dynamics\cite{Hanany-Witten}.}
Thus our assumption, that the $D_{p-2}$-branes stretched between the
final state of the $D_p$-brane are the cosmic strings in the effective
Lagrangian seems appropriate to explain the $\theta$ dependence.

\begin{figure}[ht]
 \begin{center}
\begin{picture}(300,100)(0,0)
\resizebox{10cm}{!}{\includegraphics{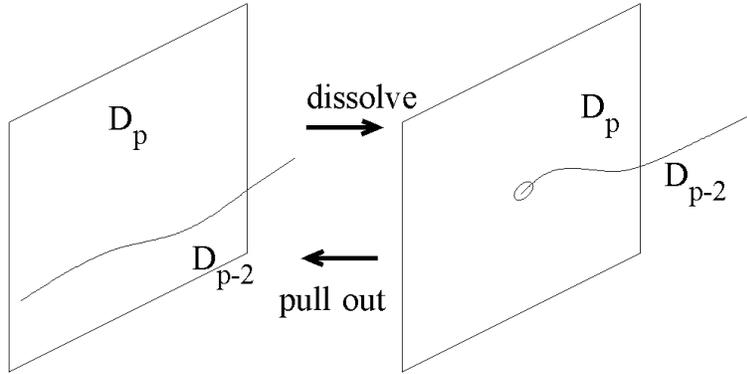}} 
\end{picture}
\caption{Bound state of $D_p$-brane and $D_{p-2}$-brane. 
In the right picture, the $D_{p-2}$-brane is dissolved in the $D_p$-brane.
The dissolved $D_{p-2}$-brane can be pulled out of the $D_p$-brane,
although the process costs energy.}  
\label{fig:dissolve}
 \end{center}
\end{figure}

\begin{figure}[ht]
 \begin{center}
\begin{picture}(410,100)(0,0)
\resizebox{15cm}{!}{\includegraphics{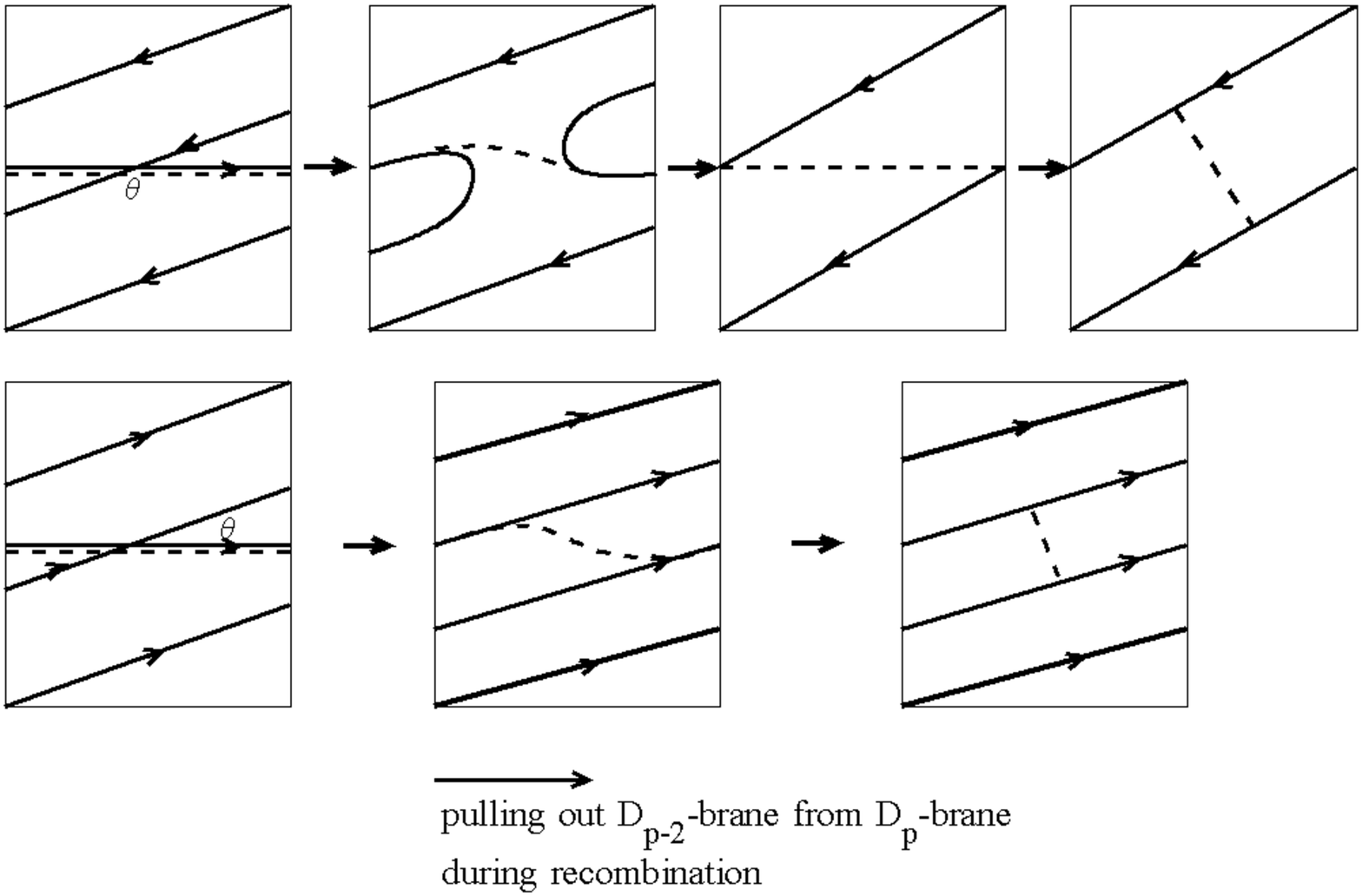}} 
\end{picture}
\caption{Upper row: schematic recombination of two $D_p$-branes with
  $(\pi-\theta) \ll 1$. The dashed line on the
  $D_p$-brane represents the $D_{p-2}$-brane that might appear on the
  worldvolume of the $D_p$-brane when the tachyon condenses.
  Second row: schematic recombination of two $D_p$-branes with $\theta \ll 1$.
  In both cases, the $D_{p-2}$-brane is pulled out of the
  $D_p$-brane.
  As a result, the daughter $D_{p-2}$ brane does not wrap the same
  compactified space as the mother brane.
  Thus, for the angled inflation model, the
  cosmic strings can depend on $\theta$, which is the property required
  from analysis of the effective Lagrangian.}
\label{fig:split}
 \end{center}
\end{figure}

However, to justify our assumption, we must answer why the tachyon
condensation on the worldvolume 
of the original $D_p$-brane could have produced such a brane configuration
during recombination after angled inflation.
To explain our idea, first we will consider a bound state of a
$D_p$-brane and a $D_{p-2}$-brane, which is depicted in fig.\ref{fig:dissolve}.
The $D_{p-2}$-brane would dissolve in the $D_p$-brane when the distance becomes
small; however, the $D_{p-2}$-brane can be pulled out from the $D_p$-branes
at the cost of energy.
Let us consider a situation where a $D_{p-2}$ brane is produced on the
worldvolume of the original $D_p$-brane due to tachyon condensation.
The initial configuration is schematically given in the left column of
fig.\ref{fig:split}. 
Then recombination starts, and the $D_{p-2}$-brane is pulled out from the
$D_p$-brane as shown in
the center of fig.\ref{fig:split}. 
The final state is given on the right of fig.\ref{fig:split},
where the dashed line denotes the $D_{p-2}$-brane that corresponds to the
cosmic string in the effective Lagrangian. 
We think one can easily understand from fig.\ref{fig:split} why the
final state of the produced 
$D_{p-2}$-branes does not wrap the same compactified space as the
original $p$-brane.
The tension of the cosmic strings should depend on $\theta$, which is
consistent with the expectation from the analysis from the effective
Lagrangian.

Now we will examine if the tension of the D-term string that is
calculated from the effective Lagrangian matches with the extended
$D_{p-2}$-brane in the brane dynamics.
Our assumption is that the cosmic string that would be produced after
angled brane inflation corresponds to the extended $D_{p-2}$-brane.
In this case, from the viewpoint of brane dynamics, the cosmic string
becomes a BPS state of the four-dimensional effective Lagrangian.
On the other hand, from the viewpoint of the effective
Lagrangian, the string that satisfies the BPS conditions must be the D-term
string, as is discussed in ref.\cite{D-term-string}.
Thus, if our assumption is correct, angled brane inflation must be a
model of D-term inflation.
Here we will consider a typical form of the Lagrangian for D-term inflation
and calculate the tension of the D-term string.
It is not a trivial task to examine if the tension of the
D-term string obtained matches with the tension that is obtained from the brane
dynamics. 
For simplicity, we will consider two $D_4$-branes on a two-dimensional torus,
which is schematically depicted in fig.\ref{fig:split}.
The radius of the torus is $r_{\|}=l_{\|}/2\pi$.
Here we consider a generic form of the potential for D-term
inflation\cite{D-terminflation}, 
\begin{equation}
\label{D-term-inflation-potential}
V= g^2\left(|\phi|^2-\xi\right)^2.
\end{equation}
Here the trigger field $\phi$ is the tachyon.
Terms that are irrelevant for our present discussions are neglected.
Thus, the energy density during inflation is represented as
\begin{equation}
\label{1st-eq}
V_0=g^2 \xi^2.
\end{equation}
One can calculate the explicit form of the energy
density $V_0$ from the brane dynamics of angled
inflation\cite{angled-defect} for $\theta \ll 1$,  
\begin{equation}
\label{2nd-eq}
V_0 = \frac{\tau_4 l_{\|} \theta^2}{4}.
\end{equation}
The tension of the D-term string in the effective Lagrangian
is\cite{D-term-string} 
\begin{equation}
\label{3rd-eq}
\mu_{D-string} =2\pi\xi,
\end{equation}
where the gauge coupling $g$ in the effective Lagrangian is given
by\cite{angled-defect} 
\begin{equation}
\label{4th-eq}
g^{-2} = \frac{M_s r_{\|}}{2\pi g_s}.
\end{equation}
From eqs.(\ref{1st-eq}), (\ref{2nd-eq}), (\ref{3rd-eq}) and
(\ref{4th-eq}), one can calculate the tension of the D-term string in
the effective Lagrangian.

On the other hand, from the pure brane dynamics, we can calculate the tension
of the cosmic string that corresponds to the
extended $D_2$-brane.
If the angle $\theta$ is small, i.e.,$\theta \ll 1$,
\begin{equation}
\mu_{D_2} = \frac{1}{(2\pi)^2}\frac{M_s^3 (l_{\|}/2)\theta}{g_s}.
\end{equation}
It is now clear that the tension of the D-term string matches
the tension calculated from the extended $D_2$-brane.

\section{Conclusions}
We have examined string production after angled brane inflation.
First, we pointed out that there has been a discrepancy
between the expected tension of the cosmic string calculated from the viewpoint
of the four-dimensional effective Lagrangian and the one obtained
from the brane tension of the daughter brane.  
If the cosmic string is a lower-dimensional daughter brane that
wraps the same compactified space as the original mother brane,
the tension of the string cannot depend on the angle.
On the other hand, from the analysis of the effective Lagrangian,
it is easy to see that the tension of the cosmic string must be
proportional to $\theta$ for $\theta \ll 1$. 
This is the problem we have solved in this paper.
Our idea is shown in fig.\ref{fig:split}.
We have calculated the explicit form of the tension of the cosmic
string, and compared the two results.

The cosmological constraint for angled inflation will be relaxed 
because the expected tension of the cosmic string becomes smaller than
the one obtained in the previous arguments by a factor of $\theta$.

\section{Acknowledgments}
We wish to thank K.Shima for encouragement, and our colleagues in
Tokyo University for their kind hospitality.

\begin{thebibliography}{1}
\bibitem{Extra_1}
I. Antoniadis, N. A-Hamed, S. Dimopoulos, and  G. R. Dvali,
{\it New dimensions at a millimeter to a fermi and superstrings at a TeV,
Phys.Lett.B436(1998)257} [hep-ph/9804398];
I. Antoniadis,
{\it A possible new dimension at a few TeV, Phys.Lett.B246(1990)377};
N. A-Hamed, S. Dimopoulos and G. R. Dvali, 
{\it The hierarchy problem and new dimensions at a millimeter,
Phys.Lett.B429(1998)263} [hep-ph/9803315].
\bibitem{low_inflation}
% added references start
N. Arkani-Hamed, S. Dimopoulos, N. Kaloper, and J. March-Russell,
{\it Rapid asymmetric inflation and early cosmology in theories with
submillimeter dimensions, Nucl.Phys.B567(2000)189} [hep-ph/9903224];
% added references end
R. N. Mohapatra, A. Perez-Lorenzana, and C. A. de S. Pires,
{\it Inflation in models with large extra dimensions driven by a bulk
scalar field, Phys.Rev.D62(2000)105030} [hep-ph/0003089];
A. Mazumdar, {\it Extra dimensions and inflation,
	Phys.Lett.B469(1999)55}[hep-ph/9902381];
A. M. Green and A. Mazumdar, 
{\it Dynamics of a large extra dimension inspired hybrid inflation model,
Phys.Rev.D65(2002)105022} [hep-ph/0201209];
D. H. Lyth, {\it Inflation with TeV scale gravity needs supersymmetry,
Phys.Lett.B448(1999)191} [hep-ph/9810320];
%added references start
P. Kanti and K. A. Olive,
{\it On the realization of assisted inflation, Phys.Rev.D60(1999)043502}
[hep-ph/9903524];
P. Kanti and K. A. Olive,
{\it Assisted chaotic inflation in higher dimensional theories,
 Phys.Lett.B464(1999)192} [hep-ph/9906331];
% added references end
T. Matsuda,
{\it Kaluza-Klein modes in hybrid inflation, Phys.Rev.D66(2002)107301}
[hep-ph/0209214];
T. Matsuda,
{\it Successful D term inflation with moduli, Phys.Lett.B423(1998)35}
[hep-ph/9705448].
\bibitem{matsuda_nontach}
T. Matsuda,
{\it Thermal hybrid inflation in brane world, Phys.Rev.D68(2003)047702}
[hep-ph/0302253]
T. Matsuda,
{\it F term, D term and hybrid brane inflation, JCAP 0311(2003)003}
[hep-ph/0302078]
T. Matsuda,
{\it Nontachyonic brane inflation, Phys.Rev.D67(2003)083519}
[hep-ph/0302035]
\bibitem{matsuda_defectinfla}
T. Matsuda,
{\it Topological hybrid inflation in brane world, JCAP 0306(2003)007}
[hep-ph/0302204]
\bibitem{low_baryo}
G. R. Dvali, G. Gabadadze, 
{\it Nonconservation of global charges in the brane universe and baryogenesis,
Phys.Lett.B460(1999)47} [hep-ph/9904221];
A. Masiero, M. Peloso, L. Sorbo, and R. Tabbash, 
{\it Baryogenesis versus proton stability in theories with extra dimensions,
Phys.Rev.D62(2000)063515} [hep-ph/0003312];
A. Pilaftsis, {\it Leptogenesis in theories with large extra dimensions,
Phys.Rev.D60(1999)105023} [hep-ph/9906265];
R. Allahverdi, K. Enqvist, A. Mazumdar and A. Perez-Lorenzana,
{\it Baryogenesis in theories with large extra spatial dimensions,
Nucl.Phys. B618(2001)377} [hep-ph/0108225];
S. Davidson, M. Losada, and A. Riotto,
{\it A new perspective on baryogenesis, Phys.Rev.Lett.84(2000)4284}
[hep-ph/0001301].
\bibitem{Defect-baryo-largeextra}
T. Matsuda,
{\it Baryon number violation, baryogenesis and defects with extra
dimensions, Phys.Rev.D66(2002)023508} [hep-ph/0204307];
T. Matsuda,
{\it Activated sphalerons and large extra dimensions,
Phys.Rev.D66(2002)047301} [hep-ph/0205331];
T. Matsuda,
{\it Enhanced baryon number violation due to cosmological defects with
localized fermions along extra dimensions, Phys.Rev.D65(2002)107302}
[hep-ph/0202258];
T. Matsuda,
{\it Defect mediated electroweak baryogenesis and hierarchy,
J.Phys.G27(2001)L103} [hep-ph/0102040].
\bibitem{Defect-baryo-4D}
T. Matsuda,
{\it Hybridized Affleck-Dine baryogenesis, Phys.Rev.D67(2003)127302}
[hep-ph/0303132];
T. Matsuda,
{\it Affleck-Dine baryogenesis after thermal brane inflation,
Phys.Rev.D65(2002)103501} [hep-ph/0202209];
T. Matsuda,
{\it Affleck-Dine baryogenesis in the local domain,
Phys.Rev.D65(2002)103502} [hep-ph/0202211];
T. Matsuda,
{\it Electroweak baryogenesis mediated by locally supersymmetry breaking
defects, Phys.Rev.D64(2001)083512} [hep-ph/0107314];
\bibitem{BraneQball}
T. Matsuda,
{\it Brane Q ball} [hep-ph/0402223].
\bibitem{MQCD_defect_matsuda}
T. Matsuda,
{\it Formation of cosmological brane defects} [hep-ph/0402232].
\bibitem{Hanany-Witten}
A. Hanany and E. Witten,
{\it Type IIB superstrings, BPS monopoles, and three-dimensional gauge
	dynamics, Nucl.Phys.B492(1997)152} [hep-th/9611230].
\bibitem{brane-inflation0}
G. R. Dvali and S. H. Henry Tye
{\it Brane inflation, Phys.Lett.B450(1999)72} [hep-ph/9812483].
\bibitem{angled-inflation}
% added references start
C. Herdeiro, S. Hirano and R. Kallosh,
{\it String theory and hybrid inflation / acceleration,
JHEP0112(2001)027} [hep-th/0110271];
K. Dasgupta, C. Herdeiro, S. Hirano and R. Kallosh,
{\it  D3 / D7 Inflationary model and  M theory,
Phys.Rev.D65(2002)126002} [hep-th/0203019],
% added references end
J. Garcia-Bellido, R. Rabadan and  F. Zamora,
{\it Inflationary scenarios from branes at angles, JHEP 0201(2002)036}
[hep-th/0112147];
\bibitem{tachyon0}
A. Sen, {\it Rolling tachyon, JHEP 0204(2002)048} [hep-th/0203211];
% added references start
A. Mazumdar, S. Panda and A. Perez-Lorenzana,
{\it Assisted inflation via tachyon condensation, Nucl.Phys.B614(2001)101}
[hep-ph/0107058].
% added references end
\bibitem{angled-defect}
N. Jones, H. Stoica, and S. H. H. Tye,
{\it Brane interaction as the origin of inflation, JHEP 0207(2002)051}
[hep-th/0203163];
S. Sarangi, S. H. H. Tye,
{\it Cosmic string production towards the end of brane inflation,
Phys.Lett.B536(2002)185} [hep-th/0204074];
L. Pogosian, S. H. H. Tye, I. Wasserman and M. Wyman,
{\it Observational constraints on cosmic string production during brane
inflation, Phys.Rev.D68(2003)023506} [hep-th/0304188];
% Added reference start
M. Gomez-Reino and I. Zavala,
{\it Recombination of intersecting D-branes and cosmological inflation,
JHEP0209(2002)020} [hep-th/0207278].
% added reference end
\bibitem{Majumdar_Davis}
M. Majumdar and A. Christine-Davis,
{\it Cosmological creation of D-branes and anti-D-branes, JHEP
0203(2002)056} [hep-th/0202148].
\bibitem{D-brane-strings}
G. Dvali and A. Vilenkin,
{\it Formation and evolution of cosmic D strings} [hep-th/0312007];
E J. Copeland, R. C. Myers and J. Polchinski,
{\it Cosmic F and D strings} [ hep-th/0312067].
\bibitem{Halyo}
E. Halyo, {\it Cosmic D term strings as wrapped D3 branes}
[hep-th/0312268].
\bibitem{BDKP-FI}
P. Binetruy, G. Dvali, R. Kallosh, A. Van Proeyen,
{\it Fayet-Iliopoulos terms in supergravity and cosmology} [hep-th/0402046]
\bibitem{Alice-string}
G. R. Dvali, I. I. Kogan and  M. A. Shifman,
{\it Topological effects in our brane world from extra dimensions,
Phys.Rev.D62(2000)106001} [hep-th/0006213].
\bibitem{incidental}
T. Matsuda,
{\it Incidental Brane Defects, JHEP 0309(2003)064} [hep-th/0309266].
\bibitem{D-term-string}
G. Dvali, R. Kallosh and A. Van Proeyen,
{\it D Term strings, JHEP 0401(2004)035} [hep-th/0312005].
\bibitem{D-terminflation}
E. Halyo, 
{\it Hybrid inflation from supergravity D terms, Phys.Lett.B387(1996)43}
[hep-ph/9606423];
P. Binetruy, G. Dvali, 
{\it D term inflation, Phys.Lett.B388(1996)241}
[hep-ph/9606342];
T. Matsuda, 
{\it Successful D term inflation with moduli, Phys.Lett.B423(1998)35}
[hep-ph/9606342].
\end{thebibliography}
\end{document}